# Formation of nickel-carbon heterofullerenes under electron irradiation

A. S. Sinitsa,[a] I. V. Lebedeva,[b] A. A. Knizhnik,[c,d] A. M. Popov,*[e] S. T. Skowron[f] and E. Bichoutskaia[f]

A way to produce new metal-carbon nanoobjects by transformation of a graphene flake with an attached transition metal cluster under electron irradiation is proposed. The transformation process is investigated by molecular dynamics simulations by the example of a graphene flake with a nickel cluster. The parameters of the nickel-carbon potential (I. V. Lebedeva et al. J. Phys. Chem. C, 2012, 116, 6572) are modified to improve description of the balance between the fullerene elastic energy and graphene edge energies in this process. The metal-carbon nanoobjects formed are found to range from heterofullerenes with a metal patch to particles consisting of closed fullerene and metal clusters linked by chemical bonds. The atomic-scale transformation mechanism is revealed by the local structure analysis. The average time of formation of nanoobjects and their lifetime under electron irradiation are estimated for experimental conditions of high-resolution transmission electron microscope (HRTEM). The sequence of images of nanostructure evolution with time during its observation by HRTEM is also modelled. Furthermore, the possibility of batch production of studied metal-carbon nanoobjects and solids based on these nanoobjects is discussed.

## Introduction

Since the discovery of fullerenes a wide set of endofullerenes[1] with atoms of non-carbon elements inside and heterofullerenes[2] with carbon atoms of the fullerene cage substituted by atoms of other elements have been synthesized. However, though transition metals have been extensively studied as catalysts for synthesis of carbon nanotubes (see Ref. 3 for review) and graphene[4], neither endofullerenes nor heterofullerenes which contain atoms of transition metals from groups V-VIII have not been synthesized yet.

Recent advances in transmission electron microscopy, first of all implementation of aberration corrections of electromagnetic lenses, have led not only to the possibility of the observation of single atom dynamics[5-7] but have also provided a powerful tool for investigation of activation of chemical reactions at atomic scale by irradiation. In particular, they have made it possible to observe a set of irradiation-induced chemical reactions between transition metals and carbon nanostructures.[8-13] This set includes hole formation in graphene assisted by Pd,[8-10] Ni,[9,10] and Cr[10] atoms and clusters, enhancement of the rate of hole formation in graphene near iron clusters,[11] formation of nanometre-sized hollow protrusions on nanotube sidewalls in the presence of catalytically active atoms of Re inserted into the nanotubes,[12] and cutting of a single-walled carbon nanotube with an Os cluster.[13]

It is well known that electron irradiation can be used to induce structural transformations of carbon nanostructures similar to those at high temperature. A bright examples of such a process is the formation of carbon nanoparticle with shell structure from amorphous nanoparticle[14,15] and the formation of double-walled nanotubes from single-walled nanotubes filled with fullerenes[16,17] both under electron irradiation[14,16] and in oven[15,17]. Both high temperature and electron irradiation treatment leads to formation of sulphur-terminated graphene nanoribbons inside carbon nanotubes filled with carbon- and sulphur-containing molecules.[18]

Recently the thermally activated transformation of a graphene flake to a fullerene catalysed by a nickel cluster was studied using molecular dynamics (MD) simulations.[19] In this process, which was predicted to occur at temperatures 700-800 K, the nickel cluster was observed to detach after the fullerene formation. However, new types of heterofullerenes with a patch made of a nickel cluster and fullerenes with nickel clusters attached from outside were found to form as intermediate structures during the transformation. Based on the similarity between irradiation-induced and thermally activated transformation processes, we consider here the possibility of obtaining these new nanoobjects by the transformation of a graphene flake with a nickel cluster attached at room temperature under the action of electron irradiation in high-resolution transmission electron microscope (HRTEM). Contrary to the thermally activated transformation in the oven, HRTEM allows visual control and, moreover, makes it possible to stop the transformation process instantly at the moment when the desired heterostructure is formed.

To demonstrate the possibility of obtaining new types of metal-carbon heterostructures we perform MD simulations of the transformation process under the action of electron irradiation simulated using the recently elaborated CompuTEM algorithm.[20,21] This algorithm takes into account annealing of the system between changes induced by irradiation in the local structure and also provides a sequence of images of sample evolution with time during its observation in HRTEM for given experimental conditions.

The considered process of electron irradiation-induced transformation of a graphene flake with a nickel cluster attached has a set of features analogous to those for processes in pure carbon systems. First of all, a closely related process is the graphene-fullerene transformation under electron irradiation.[22] However, other thermally activated and spontaneous processes should also be mentioned. The curving of a graphene surface observed in our simulations is partly similar to the buckling of a graphene layer on a substrate after its cooling[23] and theoretically predicted spontaneous warping of graphene sheets and nanoribbons.[24,25] The reconstruction of the graphene edge leading to a decrease in the number of dangling bonds also takes place at the formation of a closed cap at the open end of a carbon nanotube.[26,27] After formation of a closed fullerene shell as a result of sticking of polyyne chains or macrocycles with simultaneous structure rearrangement in laser or arc discharge synthesis of fullerenes (see Refs.28-30 and references therein) both capture[30,31] and emission[29,30] of $C_2$ molecules are possible. The emission of $C_2$ molecules has been proposed to play an essential role for the formation of abundant isomers of $C_{60}$ and $C_{70}$ fullerenes,[28-29] and of abundant isomers of metallofullerenes.[28,32] Here we also observe carbon atom loss during the transformation process. Transformation of few-layer graphene into multi-walled nanotubes at graphite sonication in presence of ferrocene aldehyde have been also observed.[33,34] Thus, we believe that the present study is important for understanding of the processes mentioned above.

Predictive atomistic modelling relies on the use of accurate interatomic potentials. The new potential for nickel-carbon systems has been recently elaborated on the basis of the first-generation bond-order Brenner potential.[19] Here the parameters of this potential are modified to improve description of the balance between the fullerene elastic energy and graphene edge energies, which is important for adequate modelling of the considered transformation process.

The paper is organized as follows. First we describe the developed potential and details of simulations. Then we present the results of MD simulations of the transformation and simulations of structure images in HRTEM. After that our conclusions are summarized and the possibility of production of proposed metal-carbon heterostructures is discussed.

## Methods

### A. Modification of Ni-C potential

Previously we have extended the first-generation bond-order Brenner potential[35] with the second set of parameters to describe nickel-carbon systems.[19] The potential designed in paper[19] reproduces the experimental data on the lattice constant, cohesive energy, elastic properties and energy of vacancy formation for fcc nickel and is qualitatively correct in predicting relative stability of nickel bulk phases. The relative energies of the nickel surfaces and the energies of formation of a nickel adatom and addimer on the (111) surface obtained by the density functional theory (DFT) calculations are well reproduced. The potential also describes the relative energies of carbon adatoms at different sites of the nickel (111) and (113) surfaces, of atoms in $C_6$ rings and in graphene on the nickel (111) surface and of carbon interstitials in the subsurface layer and in the bulk obtained by the DFT calculations.

Though the original Brenner potential[35] and its extension to nickel-carbon systems have been already successfully applied for simulations of various phenomena in carbon nanosystems,[19,20,36,37] they can be improved further. In particular, it can be expected that such processes as the graphene-fullerene transformation are sensitive to the balance of elastic and edge energy of curved graphene flakes. For this reason, we have tested the reliability of the designed Ni-C potential with respect to the description of these properties. The widely used second-generation Brenner potential[38] is also considered for comparison.

Let us first address the performance of the potentials for description of elastic energies of fullerenes. As the majority of literature data on elastic energies of small fullerenes are rather old, we have performed DFT calculations of energy differences between the $C_{60}$ and $C_{70}$ fullerenes and the periodic graphene layer per carbon atom using VASP code[39] with the generalized gradient approximation (GGA) density functional of Perdew, Burke, and Ernzerhof.[40] The basis set consists of plane waves with the maximum kinetic energy of 600 eV. The interaction of valence electrons with atomic cores is described using the projector augmented-wave method (PAW).[41] A second-order Methfessel–Paxton smearing[42] with a width of 0.1 eV is applied. The unit cell of the periodic graphene layer is considered in the optimized 4.271 Å x 2.466 Å x 20 Å model cell. Integration over the Brillouin zone is carried out using the Monkhorst-Pack method[43] with 24 x 36 x 1 k-point sampling. The calculations for the $C_{60}$ and $C_{70}$ fullerenes are performed using a single Γ point in the 20 Å x 20 Å x 20 Å model cell. The fullerene structures are geometrically optimized until the residual force acting on each atom becomes less than 0.01 eV/Å. The calculated elastic energies of the $C_{60}$ and $C_{70}$ fullerenes of 0.36 eV and 0.32 eV per carbon atom, respectively. These values are close to the energy of 0.41 eV inferred from the experimental measurements[44,45] for the $C_{60}$ fullerene (the energy of solid $C_{60}$ has been determined in Ref. 44 and the energy for sublimation into isolated $C_{60}$ molecules in Ref. 45).

To obtain the energy differences between the $C_{60}$ and $C_{70}$ fullerenes and the periodic graphene layer per carbon atom for the classical potentials, the $C_{60}$ and $C_{70}$ fullerenes are geometrically optimized using the conjugated gradient method. The periodic graphene layer is considered in the rectangular model cell comprising 10 rectangular unit cells consisting of four carbon atoms along the armchair direction and 17 unit cells along the zigzag direction. The size of the model cell is optimized to obtain the equilibrium graphene bond length. The comparison of the data obtained using the classical potentials and in our DFT calculations (Table 1) reveal that the Ni-C potential provides a good description of elastic properties of small fullerenes, while the second-generation Brenner potential tends to overestimate elastic energies.

The second property that is important for realistic modelling of the graphene-fullerene transformation is the graphene edge energy. To obtain it for the considered classical potentials we have calculated energies of zigzag and armchair graphene nanoribbons (GNRs) under periodic boundary conditions. The edge energy per unit edge length is found as

$$E_{\text{edge}} = \frac{1}{2L}\left(E_{\text{GNR}} - N\varepsilon_{\text{gr}}\right) , \qquad (1)$$

where $L$ is the length of the model cell along the nanoribbon axis, $N$ is the number of atoms in the model cell, $E_{\text{GNR}}$ is the GNR energy and $\varepsilon_{\text{gr}}$ is the energy per carbon atom in the periodic graphene layer. The same periodic layer as in the calculations of the fullerene elastic energies is considered. To find the energies of armchair and zigzag GNRs the size of the model cell is increased twice along the zigzag and armchair direction, respectively, while keeping the absolute positions of atoms the same. Then the GNR structures are geometrically optimized and the edge energies are derived according to formula (1). As seen from Table 1, the edge energies calculated using the Ni-C potential developed in our previous paper are 10 – 40% smaller than the values from literature obtained by the DFT calculations. For comparison, the data for the second-generation Brenner potential are also given and it is seen that this potential describes the graphene edge energies well.

**Table 1.** Graphene edge energies and elastic energies of the $C_{60}$ and $C_{70}$ fullerenes calculated using the original and modified versions of the Ni-C potential, the second-generation Brenner potential (Brenner II) and within DFT.

| Property | Original Ni-C | Modified Ni-C | Brenner II | DFT |
|---|---|---|---|---|
| Graphene edge energy (eV/Å) | | | | |
| Armchair | 0.83 | 1.25 | 1.09 | 1.01,[46,47] 1.20[46,48] |
| Zigzag | 0.73 | 1.50 | 1.04 | 1.18,[46] 1.39,[46,48] 1.35[47] |
| Fullerene elastic energies (eV per atom) | | | | |
| $C_{60}$ | 0.39 | 0.39 | 0.55 | 0.36[a] |
| $C_{70}$ | 0.34 | 0.34 | 0.49 | 0.32[a] |

[a]Present work.

To improve the balance of elastic and edge energies in the Ni-C potential we have slightly modified the $F$ function in the Brenner potential form.[19,35] Namely, we have set $F_{CC}(1,2,2) = -0.063$. This allows to improve the graphene edge energies, while keeping the elastic energies the same. Also as compared to the previous version of the potential, in the modified potential, we have symmetrized function $F$ to recover the same function as in the original first-generation Brenner potential.[35] To keep the correct energies of carbon structures on the nickel (111) surface we have also adjusted the parameters $a_{CCNi} = 0.115$ and $a_{CNiC} = 0.602$. The fullerene elastic energies and graphene edge energies calculated using the modified potential are given in Table 1.

As seen from Table 1, different from the previous version of the Ni-C potential[19] and second-generation Brenner potential,[38] the Ni-C potential with the small modifications of the parameters discussed above describes properly both the elastic energies of fullerenes and the graphene edge energies. However, the MD simulations using this modified potential at high temperature have revealed formation of metastable chains of alternating two-coordinated nickel and carbon atoms, which in reality should be very unstable. To supress this unphysical behaviour an additional correction has been introduced to the potential in the form of the $H$ function that was present in the original Brenner potential but was taken equal to zero in the previous version of the Ni-C potential.

The function $H$ is still taken equal to zero for carbon-carbon and nickel-nickel bonds and for the contribution of carbon atoms to the carbon-nickel bonds. Non-zero values of $H = H_{NiC}(N^C, N^{Ni})$ are assumed only for the contribution of nickel atoms having $N^{Ni} \leq 5$ nickel neighbours and $N^C \geq 1$ carbon neighbours in addition to the carbon atom participating in the nickel-carbon bond under consideration. Taking $H_{NiC}(N^C \geq 1, N^{Ni} \leq 3) = 1.3$ has allowed the fitting of the energy of nickel-carbon chains relative to bulk nickel and graphite to the value of 5.5 eV per nickel-carbon pair, which followed from our DFT calculations. The values $H_{NiC}(N^C \geq 1, N^{Ni} = 4,5) = 2.5$ have been chosen to decrease in magnitude the cohesive energy of bulk NiC carbide with the NaCl structure to the DFT value of 10.5 eV obtained in Ref. 49. These corrections have been shown sufficient to avoid formation of any of such structures in MD simulations. The full list of parameters of the modified potential can be found in the supplementary information.†

To summarize, the modification of the parameters of the Ni-C potential[19] has substantially improved the description of the elastic energies of fullerenes and the graphene edge energies. Moreover, the corrections introduced for low-coordinated nickel atoms help to avoid spurious effects in high-temperature dynamics of carbon–nickel systems. Thus, in the present form, the Ni-C potential is appropriate for modelling of the graphene-fullerene transformation and is applied here for simulations of this process at high temperatures and under electron irradiation.

### B. Details of MD simulations

The effect of electron irradiation on the graphene flake with the attached metal cluster is simulated using the recently elaborated CompuTEM algorithm.[20,21] In this algorithm, effective modelling of processes induced by electron irradiation is achieved by 1) taking into account only interactions between incident electrons and atoms leading to changes in the atomic structure (i.e. irradiation-induced events) and 2) considering annealing of the structure between these events independently at elevated temperature.

In the considered algorithm, such irradiation-induced events are described as follows:[20] 1) the nanostructure is equilibrated at a temperature corresponding to the experimental conditions in HRTEM, 2) a type of each atom of the nanostructure is determined based on the number and strength of its chemical bonds, 3) the possible minimal energy that can be transferred from an incident electron is assigned to each atom in accordance with the atom type determined at step 2, 4) a single electron-atom interaction event is introduced by giving a momentum distributed according to the standard theory of elastic electron scattering between a relativistic electron and a nucleus[20,50-53] to a random atom that is chosen based on the total probabilities of electron collisions with different atoms determined by the minimum transferred energies assigned at step 3 (such a description of electron-atom interaction events is adequate as the time of electron-atom interaction in HRTEM is considerably less than the MD time step), 5) MD run at a temperature corresponding to the experimental conditions with the duration sufficient for bond reorganisation, 6) the surrounding of the impacted atom is analysed again and if no change in the atom type or in the list of the nearest neighbours is detected as compared to step 2 within this time period (the impact is unsuccessful), the simulation cycle is repeated. However, if the system topology has changed (the impact is successful), an additional MD run of duration $t_{rel}$ at elevated temperature $T_{rel}$ is performed to describe the structural relaxation between successive electron impacts.

In our simulations, the MD runs at steps 1 and 5 are performed at a temperature of 300 K during 10 ps. This time is sufficient to capture all structural transformations induced by the electron collision. The temperature $T_{rel}$ and duration $t_{rel}$ of the additional

MD run at step 6 should be chosen so that the structure reconstruction due to irradiation-induced events is described as full as possible, while thermally induced transformations are virtually excluded. MD simulations of diffusion of vacancies towards graphene flake edges[54] imply that the average time necessary for the reconstruction of irradiation-induced defects in the flake at temperature 2500 K is about 100 ps. This indicates the minimal duration of the high-temperature MD run at step 6 of the described algorithm required for adequate modelling of electron irradiation of the graphene flake and this duration $t_{rel}$ = 100 ps is used in our simulations. To choose the temperature for the MD run at step 6 we have performed simulations of the thermally activated transformation (without irradiation) at temperatures 2100-2500 K. The results of these simulations and the choice of the temperature $T_{rel}$ on the basis of these results are presented in the next section.

At step 2 of the algorithm, the type of each carbon atom is determined based on the following information: (1) the numbers of carbon neighbours of the atom, his carbon neighbours and their neighbours, (2) presence of nearest-neighbour nickel atoms and (3) existence of non-hexagonal rings in the carbon bond network of the graphene flake to which the atom belongs. As chosen in our previous publication,[19] the minimal transferred energy is taken to be 10 eV for two-coordinated carbon atoms at flake edges and corners (here and below we refer to carbon atoms having one, two or three bonds with other carbon atoms as one-coordinated, two-coordinated and three-coordinated carbon atoms, respectively, independent of the number of bonds with nickel atoms), chains of two-coordinated carbon atoms, three-coordinated carbon atoms in non-hexagonal rings and three-coordinated carbon atoms close to the flake edges (not farther than two bonds from the edge atoms). The minimum transferred energy for the rest of three-coordinated carbon atoms in the flake interior is 15 eV. The minimal transferred energy for carbon adatoms on the metal cluster is taken as 5 eV so that it is below the adsorption energy of 6 eV for a carbon atom on the nickel (111) surface calculated using the same Ni-C potential. The same value is used for one-coordinated atoms. For these minimal transferred energies, the fraction of successful electron impacts is about 10%. This value is low enough to be sure that the majority of irradiation-induced events is taken into account.

Let us discuss the choice of the system size. Due to the random nature of electron impacts and huge thermodynamic fluctuations in nanosystems with hundreds of atoms and less, a very wide dispersion of the temporal and structural parameters of the transformation process is observed for the same number of carbon and nickel atoms in the system. Thus, a relatively small system should be considered to obtain a better statistics for the same starting system within the reasonable computational time. However, our previous studies on the nickel-assisted graphene-fullerene transformation at high temperatures demonstrated that with the decrease in the size of nickel cluster, the thermally activated desorption of the nickel cluster takes place at lower temperatures.[19] To avoid such thermally activated effects for small systems it is necessary to decrease the relaxation temperature $T_{rel}$ and, therefore, to increase the duration $t_{rel}$ of the additional MD run at step 6 of the algorithm. The increase of the duration $t_{rel}$ leads in turn to the increase of the computational time. Thus, to obtain good statistics and at the same time to adequately take into account the structure relaxation between electron impacts we choose an intermediate system size. Based on the results of our previous study,[19] the graphene flake consisting of $N_C = 96$ atoms and the nickel cluster consisting of $N_{Ni} = 13$ atoms are considered. In the beginning of the simulations, the flake has the shape of an ideal hexagon with six equal zigzag edges and the metal cluster is placed at a corner of the flake (see Figure 1a). The periodic boundary conditions are applied to the simulation cell of 20 nm x 20 nm x 20 nm size. The kinetic energy of electrons in the beam is 80 keV and the electron flux is 4.1·10$^6$ electrons/(s·nm$^2$).[20]

The considered Ni$_{13}$ cluster is so small that the nickel atoms are quite mobile even at room temperature. This equilibrium motion clearly dominates over the irradiation-induced events in the cluster. Therefore, interactions of electrons with nickel atoms are disregarded. Moreover, breaking/formation of bonds of the impacted carbon atom with nickel atoms is not considered as a prerequisite for the additional MD run at step 6 of the algorithm unless the impacted atom does not have bonds with nickel atoms at all before or after the electron collision, i.e. the atom type is changed in the result of this collision.

The in-house MD-kMC[55] (Molecular Dynamics – kinetic Monte Carlo) code is used. The modified Ni-C potential described in the previous subsection is applied. The integration time step is 0.6 fs. The temperature is maintained by the Berendsen thermostat,[56] with relaxation times of 0.1 ps, 3 ps and 0.3 ps for the MD runs at steps 1, 5 and 6 of the described algorithm, respectively. In the simulations of the thermally activated graphene-fullerene transformation, the relaxation time is 0.3 ps. To obtain the information on belonging of carbon atoms to non-hexagonal rings necessary for their classification at step 2 of the algorithm the topology of the carbon bond network of the flake is analysed on the basis of the "shortest-path" algorithm.[57] Two carbon atoms are considered as bonded if the distance between them does not exceed 1.8 Å, while for bonded carbon and nickel atoms, this maximal bond length is 2.2 Å. Because of the finite size of the simulation cell atoms and dimers that detach from the graphene flake and the metal cluster can cross the simulation cell several times and then stick back, while in the real system, they would fly away. To avoid these artificial reattachments atoms and dimers that detach from the system and do not stick back within 10 ps are removed.

C. Details of simulations of HRTEM images

In CompuTEM two main computational parts, molecular dynamics and image simulations, are linked by the experimental value of the electron flux, which allows up-scaling of the MD simulation time to the experimental time and determines the signal-to-noise ratio of the simulated images.

The total structure evolution time at experimental conditions can be expressed as a sum of the time periods between subsequent irradiation-induced events. The time period $t_{ev}$ between the events is defined as the inverse of the product of the overall cross-section $\sigma$ corresponding to all considered irradiation-induced events and the electron current density $j$

$$t_{ev} = 1/j\sigma, \qquad (1)$$

where the overall cross-section is assigned to each type $u$ of atoms as

$$\sigma = \sum_u N_u \sigma_u, \qquad (2)$$

where $\sigma_u$ is the cross-section corresponding to all considered irradiation-induced events, $N_u$ is the number of atoms of $u$-th type in the nanostructure. Equation (1) gives the rate of structure evolution under the influence of the e-beam and allows the direct comparison of the simulated process with experimentally observed dynamics. All impacts, including unsuccessful, which do not change the local atomic structure, are included in the estimation of the rate of structure evolution using equation (1).

Image simulations are produced using the Musli multislice code,[58] with thermal vibrations of the atoms taken into account using the frozen phonon approach.[59] In order to accurately reproduce the effects of anisotropic vibrations and fast conformational changes, especially important for the loosely-bound fragments observed during the structure evolution, 20 phonons are averaged to produce each image. The effect of the electron flux is then applied, assuming a one second exposition time $t_{exp}$ for each image. The intensity of each pixel is calculated as

$$I(x,y) = \text{Poisson random}\left[I_{sim}(x,y)jt_{exp}\Delta x \Delta y\right], \qquad (3)$$

where $I_{sim}(x,y)$ is the image intensity resulting from the multislice simulation, $j$ is the electron flux (taken to be $4.1 \cdot 10^6$ electrons/(s·nm$^2$), corresponding to the MD simulations described above) and $\Delta x \Delta y$ is the pixel size. The experimentally measured modulation transfer function (MTF) of a CCD camera[60] at an 80 kV accelerating voltage is applied to obtain an accurate signal to noise (SNR) ratio. The spherical aberration is 20 μm and the defocus is set to 5 nm, with a focus spread of 1.9 nm.

## Results and discussion

### A. MD simulations of thermally activated transformation

As discussed in the introduction, similar structural transformations can be often induced both by electron irradiation at moderate temperatures and by high temperature. However, in our MD simulations, we have to include a high-temperature stage to describe structure relaxation between successful electron impacts at affordable computational cost. To separate the effects of electron irradiation and high temperature the temperature $T_{rel}$ and duration $t_{rel}$ of this high-temperature stage should be chosen so that thermally induced transformations of the pristine structure are mainly excluded, i.e. the following condition should be fulfilled

$$N_{ir} t_{rel} \ll t_{th}, \qquad (4)$$

where $N_{ir}$ is the number of irradiation-induced events during the simulation of the irradiation-induced process and $t_{th}$ is the characteristic simulation time required for the thermally induced process analogous to the considered irradiation-induced process to take place at the elevated temperature $T_{rel}$. Therefore, simulations of irradiation-induced processes require prior knowledge on the thermally induced behaviour of the system.

To estimate the characteristic time of the thermally activated nickel-assisted graphene-fullerene transformation we have performed simulations of this process at temperatures 2100 K – 2500 K (Fig. 1). During the simulations, a set of parameters are monitored to determine the transformation moment and to give the possibility of comparing the effects of high temperature and electron irradiation on the process pathway. The total number of two-coordinated and one-coordinated carbon atoms $N_2$ (including atoms in carbon chains) characterizes the length of the flake edge and is used to detect the formation of the carbon cage. It is assumed that the moment of transformation $t_{th}$ corresponds to the decrease of the number $N_2$ twice $N_2(t_{th})/N_2^0 = 1/2$. The number of two-coordinated and one-coordinated carbon atoms with any non-zero number of bonds with the nickel cluster $N_{CNi}$ characterizes the contact area between the flake and cluster. Carbon atoms which are not bound by carbon-carbon bonds to the flake (i.e. are isolated from the flake) are considered as dissolved in the nickel cluster or adsorbed on the cluster surface. The number of such atoms is denoted as $N_d$. The dependences of numbers of pentagons $N_5$, hexagons $N_6$ and heptagons $N_7$ on time characterize changes in the local structure of the carbon bond network.

The transformation times obtained for the thermally activated process at different temperatures are listed in Table 2. The calculated temperature dependence of the average graphene-fullerene transformation time $\langle \tau \rangle$ can be approximated by Arrhenius equation (Fig. 2)

$$\langle \tau \rangle = \tau_0 \exp\left(\frac{E_a}{k_B T}\right), \qquad (5)$$

where $\tau_0 = 10^{-12.2 \pm 0.5}$ s is the pre-exponential factor, $E_a = 1.8 \pm 0.4$ eV is the activation energy, and $k_B$ is the Boltzmann constant. These kinetic parameters are in good agreement with the published data on the thermally activated nickel-assisted graphene-fullerene transformation obtained using the previous version of the Ni-C potential.[19] Fast desorption of the nickel cluster as compared to these previous simulations is related to destabilization of nickel-carbon bonds for low-coordinated (with 5 and less nickel neighbours) nickel atoms and, as a consequence, weaker adhesion of the considered small nickel cluster with the graphene edge.

The carbon cages formed in the result of the simulations contain structural defects, such as heptagons, octagons and other non-hexagonal rings as well as several two-coordinated atoms (Fig. 2). However, they can be referred to as "fullerenes" in the wide sense of this word, i.e. as closed carbon cages consisting mostly of three-coordinated carbon atoms. This definition and the structure of the formed carbon cages are consistent with the results of previous MD simulations.[19-21,30,36,37] Rearrangements inside the graphene-like network of three-coordinated carbon atoms have much larger barriers and are much slower than those at the flake edges.[36,37] For this reason, observation of the transformation of the obtained structures to the perfect fullerenes is not accessible for standard MD simulations.

In simulations in which the cluster does not leave the carbon cage before its closure, formation of metastable heterofullerenes with the metal cluster serving as a "patch" to the carbon cage takes place (Fig. 2i), in agreement with the previous simulations.[19] These structures are also referred to as "heterofullerenes" only in the wide sense. Their carbon cages contain structural defects and the metal atoms are not actually incorporated into the carbon network but rather stay together as a cluster (Fig. 2).

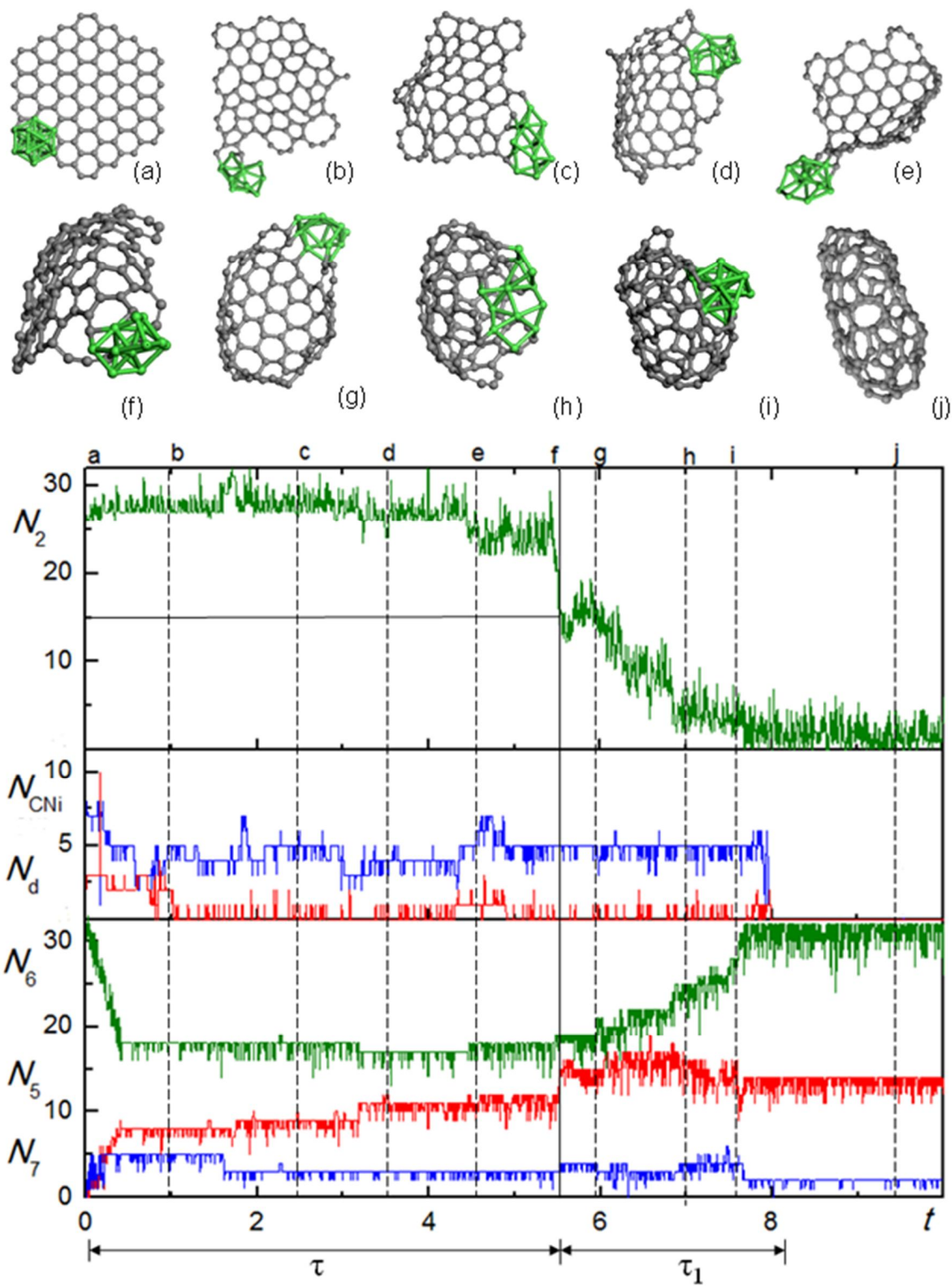

**Fig. 1** Calculated evolution of structure of the graphene flake with the attached nickel cluster at temperature 2100 K observed in: (a) 0 ns, (b) 1 ns, (c) 2.3 ns, (d) 3.5 ns, (e) 4.5 ns, (f) 5.7 ns, (g) 6 ns, (h) 7 ns, (i) 7.6 ns and (j) 9.4 ns. Calculated number $N_2$ of two-coordinated and one-coordinated carbon atoms (green line, upper panel), number $N_d$ of carbon atoms dissolved in the nickel cluster (red line, middle panel) and number $N_{CNi}$ of two-coordinated and one-coordinated carbon atoms bound to the cluster (blue line, middle panel), and numbers of pentagons $N_5$ (red line, lower panel), heptagons $N_7$ (blue line, lower panel) and hexagons $N_6$ (green line, lower panel) as functions of simulation time $t$ (in ns). The moments of time corresponding to structures (a–j) are shown using vertical dashed lines. The time $\tau$ of the graphene-heterofullerene transformation and the heterofullerene lifetime $\tau_1$ are indicated by double-headed arrows.

**Table 2.** Calculated average times $\langle \tau \rangle$ of thermally activated nickel-assisted graphene-fullerene transformation, root-mean-square deviation $\sigma$ of this time, fraction $\chi$ of simulations in which the nickel cluster detaches from the graphene flake before the transformation occurs and fraction $\xi$ of simulations in which nickel heterofullerenes are formed for different temperatures $T$. The total number $N$ of simulations for each considered system is indicated.

| $T$ (K) | $N$ | $\chi$ | $\xi$ | $\langle \tau \rangle$ (ns) | $\sigma$ (ns) |
|---|---|---|---|---|---|
| 2500 | 10 | 10/10 | 0/10 | 2.2 | 1.1 |
| 2400 | 10 | 10/10 | 0/10 | 2.6 | 1.0 |
| 2300 | 10 | 9/10 | 0/10 | 3.5 | 0.8 |
| 2200 | 10 | 8/10 | 1/10 | 7.9 | 1.8 |
| 2100 | 18 | 15/18 | 2/18 | 8.9 | 2.2 |

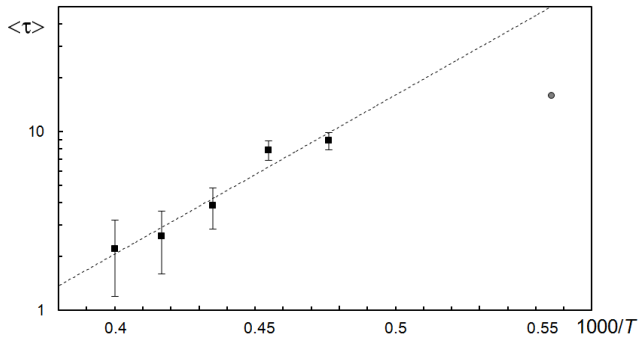

**Fig. 2.** Calculated (black squares) average time $\langle \tau \rangle$ of the thermally activated nickel-assisted graphene-fullerene transformation (in ns) and (grey circles) average total duration of the high-temperature simulation step for the irradiation-induced nickel-assisted graphene-heterostructure transformation (in ns) as functions of the reciprocal of temperature $1000/T$ (in $K^{-1}$) (logarithmic scale). The dashed line shows Arrhenius approximation (5).

## B. MD simulations of irradiation-induced transformation

As seen from the previous subsection, the graphene-fullerene transformation at temperatures above 2000 K occurs rather fast, at times of several nanoseconds. Therefore, to study the effect of electron irradiation we have to choose a lower temperature $T_{rel}$ for the high-temperature step 6 of the algorithm. Based on the 19 simulations with the temperature $T_{rel} = 1800$ K at this step, the average number of irradiation-induced events leading to the graphene-heterofullerene transformation is calculated to be $N_{ir} = 174 \pm 80$ (i.e. $N_{ir}t_{rel} = 17.4 \pm 8.0$ ns). The dependence in Fig. 2 extrapolated to the region of lower temperatures predicts that the characteristic simulation time required for the thermally induced process at this temperature is $t_{th} = 60.7$ ns. Therefore, the condition (4) is clearly fulfilled and $T_{rel} = 1800$ K is sufficient to separate the effects of electron irradiation and high temperature.

All simulation runs show that under the electron irradiation, the flat graphene flake with the nickel cluster attached finally transforms to a fullerene and the nickel cluster detaches (except 4 simulation runs which have been stopped before the cluster desorption but we are sure this should happen in continuation of these runs as well). However, contrary to thermally activated transformation the cluster always stays attached to the carbon cage until its complete closure. No formation of nickel-endofullerenes is detected, different from the case of large graphene flakes and large nickel clusters in Ref. 19. Note that the fullerene formed is too small to incorporate the cluster. In all simulation runs, the formation of metastable metal-carbon heterostructures takes place. However, the lifetimes and geometry of these metastable heterostructures are different in different simulations. Based on these qualitative differences, two limiting cases can be considered. In the first limiting case, the metal cluster is only slightly deformed during the whole simulation and keeps its compact structure (Fig. 3). The metastable heterostructures observed at the final stage of these simulations have the metal cluster linked by few nickel-carbon bonds to the carbon cage from outside. In the second limiting case, the cluster is deformed strongly during the transformation and a metastable heterofullerene with the metal cluster serving as a "patch" to the carbon cage is formed (Fig. 4).

The first stages of the transformation are similar in all cases and correspond to the stages of the irradiation-induced transformation of the graphene flake without any metal cluster attached.[20] Most structural rearrangements of the graphene flake take place at the edges and corners. The very first steps of the simulations reveal the formation of non-hexagonal polygons at the flake edges (Fig. 3b and Fig. 4b). Incorporation of pentagons at the edges leads to curving of the flake (Fig. 3c-e and Fig. 4c-e). At some moment the reconstruction of the flake edges results in the formation of a bowl-shaped region (Fig. 3f and Fig. 4f). Analogously to simulations[20,21] of the graphene-fullerene transformation in HRTEM without any cluster attached, we confirm the suggestion[22] that generation of pentagons inside the graphene flake is necessary for formation of bowl-shaped structure of the flake at the intermediate stage of the transformation. It should be also noted that many events induced by irradiation in the graphene flake correspond to formation of short chains of two-coordinated carbon atoms that are bound to the flake at one end or at both ends. Sticking of these chains back to the flake (at the free end or at one of the atoms in the middle) in the configuration different from that before the electron impact leads to reconstruction of the flake edge and generation of non-hexagonal polygons, including pentagons. Therefore, formation of such chains is very important for the transformation of the graphene flake. The same mechanism was observed in the simulations of high-temperature transformation of the pristine graphene flake[19,36,37] and thermally activated closing of carbon nanotube ends using the density-functional tight-binding method.[27]

As soon as a bowl-shaped region is formed, diverse behaviour of the system is possible. In the first limiting case, formation of carbon-carbon bonds occurs preferentially over increasing the number of carbon-nickel bonds (Fig. 3g-i). Fast zipping of the flake edges occurs, while the number $N_{CNi}$ of two-coordinated and one-coordinated carbon atoms bonded to the nickel atom decreases monotonically. As mentioned above, the metal cluster finally desorbs and the carbon cage closes completely (Fig. 3j). In the second limiting case, the increase in the number of nickel-carbon bonds occurs rather than formation of new carbon-carbon bonds, which is seen by the increase in the number $N_{CNi}$ of two-coordinated and one-coordinated carbon atoms bonded to the cluster (Fig. 4). This is achieved by smearing the nickel cluster over the edges of the carbon cage (Fig. 4g-h). As a result, a long-living metastable heterofullerene consisting of the carbon cage with the metal patch is formed (Fig. 4h,i). Nevertheless, finally, the same as in the first case, the carbon cage closes completely (Fig. 4j) and the metal cluster detaches.

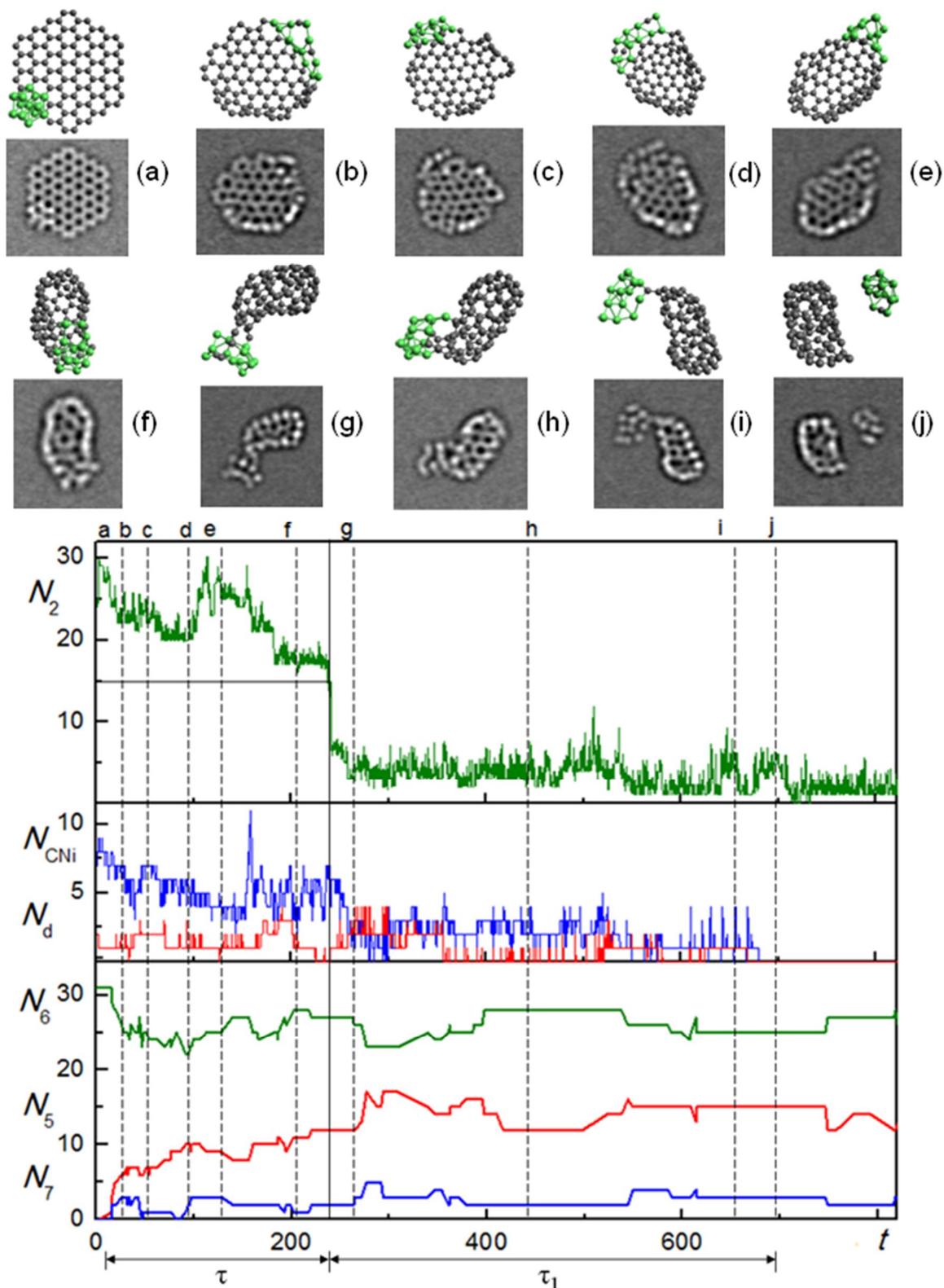

**Fig. 3** Calculated evolution of structure of the graphene flake with attached nickel cluster under electron irradiation in HRTEM observed in: (a) 0 s, (b) 33 s, (c) 54 s, (d) 95 s, (e) 117 s, (f) 201 s, (g) 267 s, (h) 438 s, (i) 670 s and (j) 695 s. Calculated number $N_2$ of two-coordinated and one-coordinated carbon atoms (green line, upper panel), number $N_d$ of carbon atoms dissolved in the nickel cluster (red line, middle panel) and number $N_{CNi}$ of two-coordinated and one-coordinated carbon atoms bound to the cluster (blue line, middle panel), and numbers of pentagons $N_5$ (red line, lower panel), heptagons $N_7$ (blue line, lower panel) and hexagons $N_6$ (green line, lower panel) as functions of time $t$ (in s) under electron irradiation in HRTEM. The moments of time corresponding to structures (a–j) are shown using vertical dashed lines. The time $\tau$ of the graphene-heterofullerene transformation and the heterofullerene lifetime $\tau_1$ are indicated by double-headed arrows

It should be noted that this pathway is qualitatively close to the pathway for the thermally activated nickel-assisted graphene-fullerene transformation (Fig. 1). Thus, the graphene-fullerene transformation is an example of the process that happens in similar ways at high temperature and under the electron irradiation.

The closed carbon cages formed under electron irradiation contain large non-hexagonal rings and two-coordinated carbon atoms (Fig. 3j and Fig. 4j), the same as in the previous simulations for a pristine graphene flake[20] and the thermally activated process.[19,21,36,37] The dynamics of non-hexagonal rings during the two discussed examples of the flake evolution are shown in Fig. 3 and Fig. 4. Similar to the previous work on the thermally activated nickel-assisted graphene-fullerene transformation[19] the metal cluster stays intact during the transformation and single nickel atoms do not incorporate into the graphene-like network of three-coordinated carbon atoms (Fig. 3 and Fig. 4). Therefore, the structures obtained in the simulations should be also considered as "fullerenes" and "heterofullerenes" in the wide sense.

The average graphene-heterofullerene transformation time under electron irradiation in all the simulation runs performed is 495 ± 200 s. This is only slightly smaller than the graphene-fullerene transformation time 630 ± 120 s obtained for the graphene flake without any cluster attached for the same conditions of observation in HRTEM in our previous publication.[20] Therefore, the metal cluster has only a mild effect on the graphene folding kinetics. Indeed the statistical data on the frequencies of irradiation-induced events for different types of carbon atoms demonstrate that 78% of these events are related to electron impacts with carbon atoms not bonded to the cluster (Table 3). It is important to note that the graphene-fullerene transformation was observed in HRTEM in the absence of a metal cluster.[22] Thus, the conclusion on the mild effect of the cluster on the transformation process is consistent with this experimental result.

Nevertheless, some influence of the cluster can be explained by the following effects. Firstly, the cluster favors wrapping the graphene flake around it as this corresponds to an increase in the number of nickel-carbon bonds. Therefore, the cluster stabilizes structural rearrangements that are directed at folding the graphene flake. The attempts of the flake to wrap around the cluster can be seen by peaks in the number $N_{CNi}$ of two-coordinated and one-coordinated carbon atoms bonded to the cluster before the bowl-shaped region is formed (Fig. 3 and 4). Secondly, the cluster introduces new irradiation-induced reactions that contribute to the transformation. For example, electron collisions can lead to the transfer of carbon atoms from the graphene flake to the cluster that can later reattach to the flake. The existence of such a process, which is impossible for the pristine graphene flake, can be noticed by the elevated number $N_d$ of carbon atoms dissolved or adsorbed on the cluster (Fig. 3 and Fig. 4) as compared to the equilibrium level for the same system at high temperature (Fig. 2). Finally, the cluster affects activation energies for reactions in the contact area between the graphene flake and the cluster, leading to changes in the cross-sections for irradiation-induced processes in that area. This effect can be illustrated by comparison of atom emission events for the pristine graphene flake[20] and the flake with the cluster attached.

**Table 3.** Calculated relative frequencies of all irradiation-induced events and emission events for impacted carbon atoms of different types.

| | Types of atoms | All events | Emission |
|---|---|---|---|
| | Not bonded to the cluster | | |
| 1. | One-coordinated atoms | 0.0167 | 0.1224 |
| 2. | Two-coordinated atoms except atoms in chains | 0.0471 | 0.0612 |
| 3. | Two-coordinated atoms in chains (having only two bonds with two-coordinated or one-coordinated carbon atoms) | 0.2012 | 0.0204 |
| 4. | Three-coordinated atoms in non-hexagonal rings | 0.3876 | 0.0204 |
| 5. | Three-coordinated carbon atoms in hexagons located not farther than by two bonds to atoms of types 1-4 and 8-11 | 0.0550 | 0.0816 |
| 6. | Edge three-coordinated atoms in hexagons (bonded to at least one two-coordinated atom) | 0.0756 | 0 |
| 7. | Other three-coordinated atoms | 0 | 0 |
| | Total: | 0.7831 | 0.3061 |
| | Bonded to the cluster | | |
| 8. | One-coordinated atoms | 0.0393 | 0.0204 |
| 9. | Two-coordinated atoms except atoms in chains | 0.0245 | 0.0204 |
| 10. | Two-coordinated atoms in chains (having only two bonds with two-coordinated or one-coordinated carbon atoms) | 0.0726 | 0.1020 |
| 11. | Three-coordinated atoms in non-hexagonal rings | 0.0020 | 0 |
| 12. | Three-coordinated carbon atoms in hexagons located not farther than by two bonds to atoms of types 1-4 and 8-11 | 0.0020 | 0 |
| 13. | Edge three-coordinated atoms in hexagons (bonded to at least one two-coordinated atom) | 0 | 0 |
| 14. | Other three-coordinated atoms | 0 | 0 |
| 15. | Adatoms | 0.0765 | 0.5510 |
| 16. | Ad-dimers | 0 | 0 |

The loss of several carbon atoms is observed in the simulations. The distribution of sizes of fullerenes formed from the graphene flake has an average of 84 atoms and a root-mean-square deviation of 5 atoms, i.e. on average 12 atoms are lost before the carbon cage closes. This is much smaller than the size of the flake, therefore, the decrease in the flake size can be disregarded in the calculations of the transformation time. On the other hand, this is much greater than the atom losses of 1-2 atoms observed in our previous work for the graphene flake without any cluster attached.[20] The statistics on the electron-induced events for impacted carbon atoms of different type reveals that in 69% of cases, the atom emission occurs for the carbon atoms bonded to the nickel cluster (Table 3). 55% of such events are emission of carbon adatoms from the nickel cluster. Thus, it can be concluded that one of the important channels for carbon atom loss is irradiation-induced transfer of carbon atoms from the graphene flake to the cluster followed by their emission from the cluster. Therefore, the cluster reduces the threshold energy for emission of carbon atoms and favours atom losses.

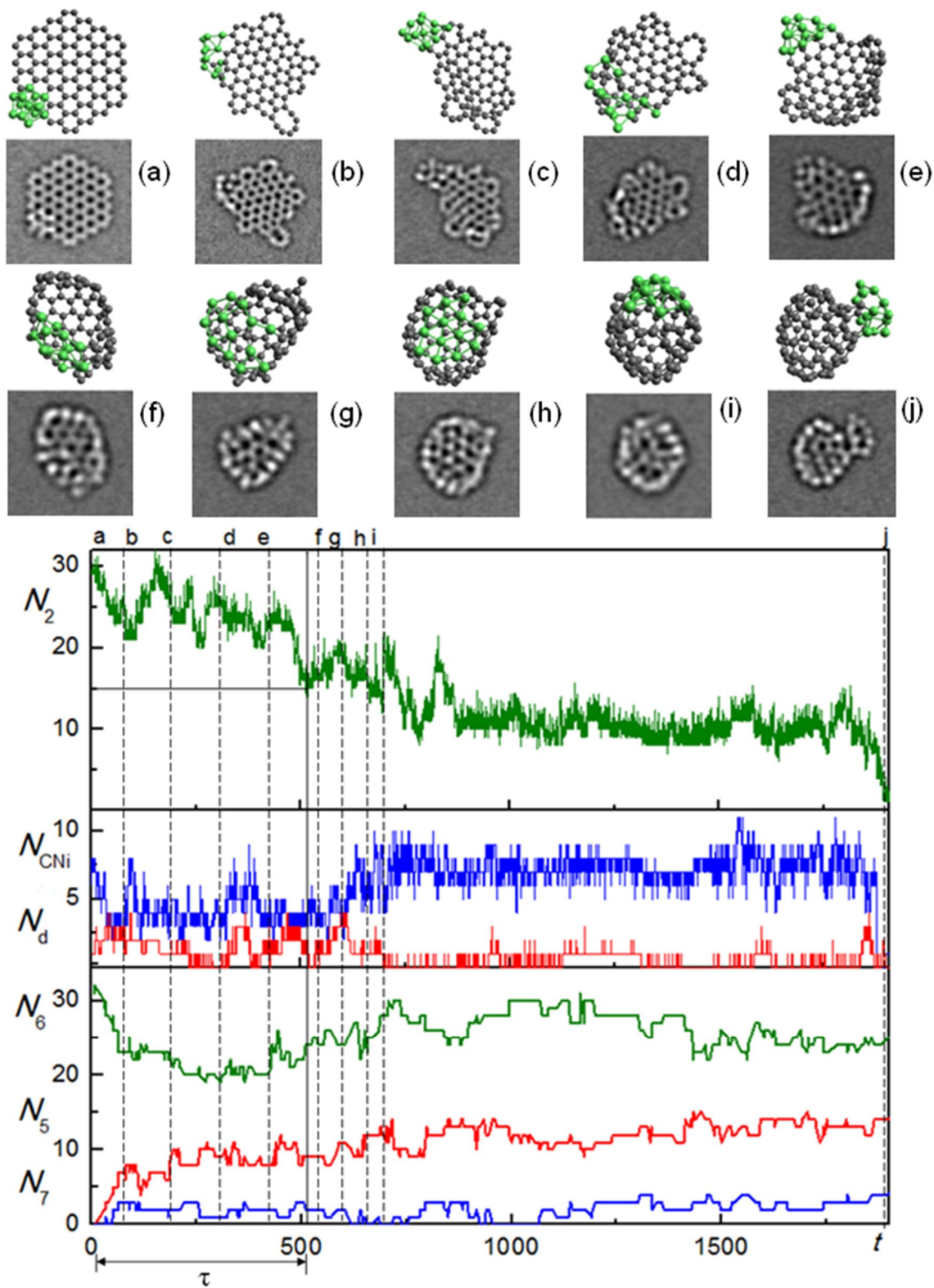

**Fig. 4** Calculated evolution of structure of the graphene flake with attached nickel cluster under electron irradiation in HRTEM observed in: (a) 0 s, (b) 113 s, (c) 207 s, (d) 306 s, (e) 461 s, (f) 550 s, (g) 632 s, (h) 689 s, (i) 704 s and (j) 1878 s. Calculated number $N_2$ of two-coordinated and one-coordinated carbon atoms (green line, upper panel), number $N_d$ of carbon atoms dissolved in the nickel cluster (red line, middle panel) and number $N_{CNi}$ of two-coordinated and one-coordinated carbon atoms bound to the cluster (blue line, middle panel), and numbers of pentagons $N_5$ (red line, lower panel), heptagons $N_7$ (blue line, lower panel) and hexagons $N_6$ (green line, lower panel) as functions of time $t$ (in s) under electron irradiation in HRTEM. The moments of time corresponding to structures (a–j) are shown using vertical dashed lines. The time $\tau$ of graphene-heterofullerene transformation is indicated by double-headed arrow.

Some conclusions on the role of the cluster in the graphene-heterofullerene transformation can be also drawn by the detailed analysis of the number $N_{CNi}$ of two-coordinated and one-coordinated carbon atoms bonded to the cluster. This quantity characterizes the contact area between the carbon structure and the cluster and, thus, reflects the degree to which the cluster participates in the transformation processes. Fig. 5a demonstrates that there is a correlation between the number $\langle N_{CNi} \rangle_{t<\tau}$ of carbon atoms bonded to the cluster averaged from the start of the simulations to the moment of graphene folding and the graphene-heterofullerene transformation time $\tau$. The general trend is that greater values of $\langle N_{CNi} \rangle_{t<\tau}$ correspond to smaller transformation times. This means that the more bonds there are between the cluster and the flake, i.e. the stronger they interact, the transformation happens faster. The transformation times range from 300 s for the clusters bonded on average to 6 two-coordinated and one-coordinated carbon atoms up to 1000 s for the clusters bonded with 4-5 two-coordinated and one-coordinated carbon atoms.

The number $N_{CNi}$ of two-coordinated and one-coordinated carbon atoms bonded to the cluster after formation of a heterostructure also correlates with its stability. To characterize the latter we determine the heterostructure lifetime $\tau_1$ as the time period between the moment $\tau$ of graphene folding to the moment of cluster desorption when $N_{CNi}$ goes to zero. For 4 simulation runs that have been stopped before the cluster desorption, the lifetime is determined as the time period between the moment $\tau$ of graphene folding and the end of the simulations. Fig. 5b shows that this lifetime increases with increasing the number of $\langle N_{CNi} \rangle_{t>\tau}$ averaged over this time period. This can be interpreted in the way that stronger interaction between the cluster and carbon cage provides more stable heterostructures. The lifetime varies from less than 50 s in the case when the cluster is bonded to only 2-3 two-coordinated and one-coordinated carbon atoms (heterostructures with the cluster linked by few nickel-carbon bonds to the carbon cage from outside, such as in Fig. 3) to 1000 s in the case when the cluster is bonded to more than 6 two-coordinated and one-coordinated carbon atoms (heterofullerenes with the cluster serving as a metal patch to the carbon cage, such as in Fig. 4). However, in any case this lifetime is sufficient for stopping electron irradiation in HRTEM at the moment when the heterostructure exists. Examples of images that can be observed in HRTEM during the graphene-heterofullerene transformation presented in Fig. 3 and 4 show that distinguishing between the considered types of metal-carbon heterostructures is possible under a visual control in HRTEM.

Moreover, the lifetimes of the heterostructures formed in the result of the graphene transformation are on the order of the transformation time. The time dependence of the number $N_h$ of heterostructures that are in the transformation stage between the moments of graphene folding and cluster desorption has a maximum at 700 s (Fig. 6) with the yield greater 50 % during 500 s. Therefore, such heterostructures can be synthesized experimentally with a high yield without any visual control in HRTEM if the electron irradiation is stopped at the moment, when the maximum yield of the heterostructures is achieved.

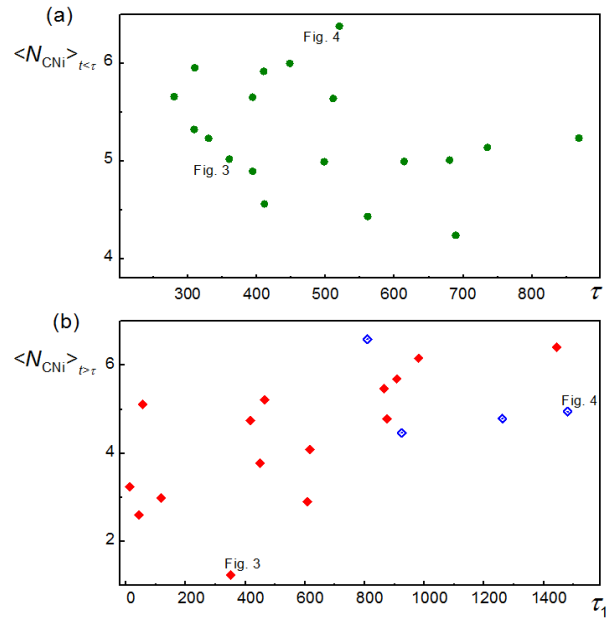

**Fig. 5.** (a) Calculated number $\langle N_{CNi} \rangle_{t<\tau}$ of two-coordinated and one-coordinated carbon atoms bonded to the cluster averaged from the start of the simulations to the moment of graphene folding as a function of the graphene-heterofullerene transformation time $\tau$ (in s) under electron irradiation in HRTEM. (b) Calculated number $\langle N_{CNi} \rangle_{t>\tau}$ of two-coordinated and one-coordinated carbon atoms bonded to the cluster averaged from the moment of graphene folding to the moment of cluster desorption (filled symbols) or to the end of simulations (open symbols) as a function of the heterofullerene lifetime $\tau_1$ (in s) under electron irradiation in HRTEM. The data corresponding to evolution of structure presented in Figs. 3 and 4 are indicated.

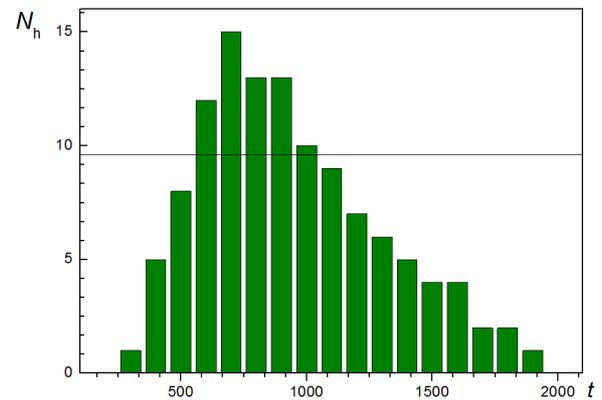

**Fig. 6.** Calculated number $N_h$ of heterostructures in the transformation stage between the moments of graphene folding and cluster desorption for 19 simulation runs at different moments of time $t$ (in s) under electron irradiation in HRTEM. The horizontal line corresponds to the 50% yield of heterostructures.

## Conclusions

Based on the DFT calculations, the parameters of the Ni-C

potential[19] have been modified to improve the description of the balance of elastic and edge energy of curved graphene flakes. Moreover, additional corrections have been introduced to supress spurious dynamic effects that can appear at high temperatures, such as formation of chains of alternating two-coordinated nickel and carbon atoms. The corrected potential allows to perform more realistic simulations of pure carbon and nickel-carbon systems.

The potential has been applied for the MD simulations of the transformation of the graphene flake consisting of 96 atoms with the $Ni_{13}$ cluster attached both at high temperature and under the action of electron irradiation in HRTEM. Based on these simulations, the temperature 1800 K is chosen as appropriate for modelling of structure relaxation between successive electron impacts. The first stages of the irradiation-induced transformation of the graphene flake with the attached nickel cluster are shown to be the same as for the pristine graphene flake,[20] with non-hexagonal rings appearing rapidly at the flake edges. However, upon formation of the bowl-shaped region, the system behaviour shows diversity. In first limiting case with the preferential formation of nickel-carbon bonds over carbon-carbon bonds, the transformation to the heterofullerene with the nickel cluster serving as a patch to the carbon cage is observed. In the opposite case, the heterostructure with the metal cluster linked to the nearly closed carbon cage just by a couple of bonds and hanging outside of this cage is formed. This pathway of the irradiation induced graphene-heterofullerene transformation is qualitatively close to that for the thermally activated transformation, which confirms the similarity of the thermally activated and irradiation-induced processes for carbon nanostructures.

It is shown that the presence of the cluster has a mild effect on the overall kinetics of the irradiation-induced transformation. The role of the cluster in this process is related to (1) wrapping the graphene flake around it, (2) introduction of new irradiation-induced events, and (3) changing threshold energies of irradiation-induced events. In particular, it is found that the cluster reduces the threshold energy for emission of carbon atoms and favours atom losses. Therefore, the combined effect of the nickel cluster and electron irradiation can be used for controlled cutting of carbon nanostructures. We propose that this effect can explain the observed in HRTEM cutting of nanotubes by Os cluster.[13]

The correlations between the number of carbon atoms bonded to the cluster, transformation time and lifetime of the heterostructure formed are revealed. These correlations demonstrate that stronger interactions between the cluster and carbon structure speed up the transformation process and stabilize the heterostructure. The calculated transformation time and heterostructure lifetimes are on the order of hundreds of seconds suggesting that all stages of the transformation can be resolved and controlled in HRTEM. The sequences of images of structure evolution under observation in HRTEM are obtained for the two limiting cases of the transformation mechanism. The maximum yield of 80% of heterostructures is predicted for 700 s of observation.

Recent advances in technologies of manipulation with individual nanoobjects makes us expect that the considered process of controlled synthesis of new types of metal-carbon heterosctructures in HRTEM can be implemented experimentally in the near future. Pioneering works[61,62] of Eigler have been followed by significant developments in controlled manipulation of atoms and molecules on surfaces (see reviews for atom/molecule manipulation using scanning tunneling microscopy[63] and atomic force microscopy[64]). In particular, it should be pointed out that the methods to cut a graphene layer into flakes of certain size and shape have been elaborated.[65] Assembly and disassembly of metal clusters with a precise control of single atoms has been also demonstrated.[66] Thus, using present-day technologies, it is already possible to prepare metal clusters of a controlled size and composition (atoms of different elements can be combined into one cluster) attached to graphene flakes of a controlled size and shape. After that we suggest to perform the transformation similar to the graphene-fullerene transformation, which was already demonstrated experimentally in HRTEM,[22] but with an added nickel cluster. Therefore, there are no significant obstacles that could impede experimental observation of the studied process in HRTEM.

Moreover, methods to produce a 2D network from identical large flat organic molecules on a surface have been developed.[67,68] Such a surface network made of large polycyclic hydrocarbons with transition metal clusters deposited on it can be used as a basis for batch production of the proposed heterostructures (hydrogen is removed rapidly by electron irradiation). Since the time period during which the yield of the heterostructures is greater than 50% is the same as the average transformation time we believe that methods of batch production of the proposed metal-carbon heterostructures and new types of 2D and 3D metal-organic structures made of them can be implemented.

## Acknowledgements

AS, AK and AP acknowledges Russian Foundation of Basic Research (11-02-00604-a). AP acknowledges Samsung Global Research Outreach Program. IL acknowledges support by the Marie Curie International Incoming Fellowship within the 7th European Community Framework Programme (Grant Agreement PIIF-GA-2012-326435 RespSpatDisp), Grupos Consolidados del Gobierno Vasco (IT-578-13) and the computational time on the Supercomputing Center of Lomonosov Moscow State University[69] and the Multipurpose Computing Complex NRC "Kurchatov Institute."[70] EB acknowledges EPSRC Career Acceleration Fellowship, New Directions for EPSRC Research Leaders Award (EP/G005060), and ERC Starting Grant for financial support.

## Notes and references

*Corresponding author: Andrey Popov  E-mail: popov-isan@mail.ru*
*[a] Moscow Institute of Physics and Technology, Institutskii pereulok 9, Dolgoprudny 141700, Moscow Region, Russia.*
*[b] Nano-Bio Spectroscopy Group and ETSF Scientific Development Centre, Departamento de Fisica de Materiales, Universidad del Pais Vasco UPV/EHU, San Sebastian E-20018, Spain.*
*[c] Kintech Lab Ltd., Kurchatov Square 1, Moscow 123182, Russia.*
*[d] National Research Centre "Kurchatov Institute", Kurchatov Square 1, Moscow 123182, Russia.*
*[e] Institute of Spectroscopy of Russian Academy of Sciences, Fizicheskaya Street 5, Troitsk, Moscow 142190, Russia.*
*[f] Department of Chemistry, University of Nottingham, University Park, Nottingham NG7 2RD, United Kingdom.*

*†Supplementary Information available: The full list of the parameters of the modified Ni-C potential.*

1. H. Shinohara, *Rep. Prog. Phys.*, 2000, **63**, 843.
2. O. Vostrowsky and A. Hirch, *Chem. Rev.*, 2006, **106**, 5191.
3. V. Jourdain and C. Bichara, *Carbon*, 2013, **58**, 2.
4. K. S. Kim, Y. Zhao, H. Jang, S. Y. Lee, J. M. Kim, K. S. Kim, J.-H. Ahn, P. Kim, J.-Y. Choi and B. H. Hong, *Nature* 2009, **457**, 706.
5. J. C. Meyer, C. O. Girit, M. F. Crommie and A. Zettl, *Nature*, 2008, **454**, 319.
6. J. C. Meyer, C. Kisielowski, R. Erni, M. D. Rossell, M. F. Crommie and A. Zettl, *Nano Lett.*, 2008, **8**, 3582.
7. R. Erni, M. D. Rossell, P. Hartel, N. Alem, K. Erickson, W. Gannett and A. Zettl, *Phys. Rev. B*, 2010, **82**, 165443.
8. R. Zan, Q. Ramasse, U. Bangert and K. S. Novoselov, *Nano Lett.*, 2012, 12, 3936.
9. Q. M. Ramasse, R. Zan, U. Bangert, D. W. Boukhvalov, Y.-W. Son and K. S. Novoselov, *ACS Nano*, 2012, **6**, 4063.
10. R. Zan, U. Bangert, Q. Ramasse and K. S. Novoselov, *J. Phys. Chem. Lett.* 2012, **3**, 953.
11. J. Campos-Delgado, D. L. Baptista, M. Fuentes-Cabrera, B. G. Sumpter, V. Meunier, H. Terrones, Y. A. Kim, H. Muramatsu, T. Hayashi, M. Endo, M. Terrones and C. A. Achete, *Part. Part. Syst. Charact.*, 2013, **30** 76.
12. T. W. Chamberlain, J. C. Meyer, J. Biskupek, J. Leschner, A. Santana, N. A. Besley, E. Bichoutskaia, U. Kaiser and A. N. Khlobystov, *Nat. Chem.*, 2011, **3**, 732.
13. T. Zoberbier, T. W. Chamberlain, J. Biskupek, N. Kuganathan, S. Eyhusen, E. Bichoutskaia, U. Kaiser and A. N. Khlobystov, *J. Am. Chem. Soc.*, 2012, **134**, 3073.
14. D. Ugarte, *Nature*, 1992, **359**, 707.
15. D. Ugarte, *Chem. Phys. Lett.*, 1993, **207**, 473.
16. J. Sloan, R. E. Dunin-Borkowski, J. L. Hutchison, K. S. Coleman, V. C. Williams, J. B. Claridge, A. P. E. York, C. G. Xu, S. R. Bailey, G. Brown, S. Friedrichs, M. L. H. Green, *Chem. Phys. Lett.*, 2000, **316**, 191.
17. S. Bandow, M. Takizawa, K. Hirahara, M. Yadasako and S. Iijima, *Chem. Phys. Lett.*, 2001, **337**, 48.
18. T. W. Chamberlain, J. Biskupek, G. A. Rance, A. Chuvilin, T. J. Alexander, E. Bichoutskaia, U. Kaiser and A. N. Khlobystov, *ACS Nano*, 2012, **6**, 3943.
19. I. V. Lebedeva, A. A. Knizhnik, A. M. Popov and B. V. Potapkin, *J Phys. Chem. C*, 2012, **116**, 6572.
20. S. T. Skowron, I. V. Lebedeva, A. M. Popov and E. Bichoutskaia, *Nanoscale*, 2013, **5**, 6677.
21. A. Santana, A. Zobelli, J. Kotakoski, A. Chuvilin and E. Bichoutskaia, *Phys. Rev. B*, 2013, **87**, 094110.
22. A. Chuvilin, U. Kaiser, E. Bichoutskaia, N. A. Besley and A. N. Khlobystov, Nat. Chem., 2010, 2, 450.
23. Z. Li, Z. Cheng, R. Wang, Q. Li, and Y. Fang, *Nano Lett.*, 2009, **9**, 3599.
24. V. B. Shenoy, C. D. Reddy, A. Ramasubramanian and Y.-W. Zhang, *Phys. Rev. Lett.* 2008, **101**, 245501.
25. V. B. Shenoy, C. D. Reddy, and Y.-W. Zhang, *ACS Nano*, 2010, **4**, 4840.
26. S. Irle, G. Zheng, M. Elstner, and K. Morokuma, *Nano Lett.*, 2003, **3**, 465.
27. G. Zheng, S. Irle, M. Elstner, and K. Morokuma, *J. Phys. Chem. A*, 2004, **108**, 3182.
28. Y.E. Lozovik and A.M. Popov, *Usp. Fiz. Nauk*, 1997, **167**, 751.
29. S. Irle, G. Zheng, Z. Wang, and K. Morokuma, *J. Phys. Chem. B*, 2006, **110**, 14531.
30. B. Saha, S. Irle, and K. Morokuma, *J. Phys. Chem. C*, 2011, **115**, 22707.
31. P. W. Dunk, N. K. Kaiser, C. L. Hendrickson, J. P. Quinn, C. P. Ewels, Y. Nakanishi, Y. Sasaki, H. Shinohara, A. G. Marshall and H. W. Kroto, *Nat. Commun.,* 2012, **3**, 855.
32. J. Zhang, F. L. Bowles, D. W. Bearden, W. K. Ray, T. Fuhrer, Y. Ye, C. Dixon, K. Harich, R. F. Helm, M. M. Olmstead, A. L. Balch, and H. C. Dorn, *Nat. Chem.*, 2013, **5**, 880.
33. M. Quintana, M. Grzelczak, K. Spyrou, M. Calvaresi, S. Bals, B. Kooi, G. Van Tendeloo, P. Rudolf, F. Zerbetto, and M. Prato, *J. Am. Chem. Soc.*, 2012, **134**, 13310.
34. M. Calvaresi, M. Quintana, P. Rudolf, F. Zerbetto, and M. Prato, *Chem. Phys. Chem.*, 2013, **14**, 3437.
35. D. W. Brenner, *Phys. Rev B.* 1990, **42**, 9458.
36. I. V. Lebedeva, A. A. Knizhnik, A. A. Bagatur'yants and B. V. Potapkin, *Physica E*, 2008, **40**, 2589.
37. I. V. Lebedeva, A. A. Knizhnik and B. V. Potapkin, *Russian Journal of Physical Chemistry B*, 2007, **26**, 675.
38. D. W. Brenner, O. A Shenderova, J. A. Harrison, S. J. Stuart, B. Ni and S. B. Sinnott, *Phys.: Condens. Matter*, 2002, **14**, 783.
39. G. Kresse and J. Furthmüller, *Phys. Rev. B*, 1996, **54**, 11169.
40. J. P. Perdew, K. Burke and M. Ernzerhof, *Phys. Rev. Lett.*, 1996, **77**, 3865.
41. G. Kresse and D. Joubert, *Phys. Rev. B*, 1999, **59**, 1758.
42. M. Methfessel and A. T. Paxton, *Phys. Rev. B*, 1989, **40**, 3616.
43. H. J. Monkhorst and J. D. Pack, *Phys. Rev. B*, 1976, **13**, 5188.
44. H. D. Beckhaus, C. Ruchardt, M. Kao, F. Diederich and C. S. Foote, *Angew. Chem.*, 1992, **31**, 63.
45. C. Pan, M. P. Sampson, Y. Chai, R. H. Hauge and J. L. Margrave, *J. Phys. Chem.*, 1991, **95**, 2944.
46. Y. Liu, A. Dobrinsky and B. I. Yakobson, *Phys. Rev. Lett.*, 2010, **105**, 235502.
47. J. Gao, J. Yip, J. Zhao, B. I. Yakobson and F. Ding, *J. Am. Chem. Soc.*, 2011, **133**, 5009.
48. C. K. Gan and D. J. Srolovitz, *Phys. Rev. B*, 2010, **81**, 125445.
49. W. Xiao, M. I. Baskes and K. Cho, *Surf. Sci.*, 2009, **603**, 1985.
50. N. Mott, *Proc. R. Soc. London, Ser. A*, 1929, **124**, 426.
51. N. Mott, *Proc. R. Soc. London, Ser. A*, 1932, **135**, 429.
52. W. McKinley and H. Feshbach, *Phys. Rev.*, 1948, **74**, 1759.
53. A. Zobelli, A. Gloter, C. P. Ewels, G. Seifert and C. Colliex, *Phys. Rev. B*, 2007, **75**, 245402.
54. A. Santana, A. M. Popov and E. Bichoutskaia, *Chem. Phys. Lett.,* 2013, **557**, 80.
55. http://www.kintechlab.com/products/md-kmc/
56. H. J. C. Berendsen, J. P. M. Postma, W. F. van Gunsteren, A. DiNola and J. R. Haak, *J. Chem. Phys.*, 1984, **81**, 3684.
57. D. C. Franzblau, *Phys. Rev. B*, 1991, **44**, 4925.
58. A. Chuvilin and U. Kaiser, *Ultramicroscopy*, 2005, **104**, 73.
59. Z. L.Wang, Elastic and Inelastic Scattering in Electron Diffraction and Imaging (Plenum Press, New York, 1995).
60. A. Lubk, F. Roder, T. Niermann, C. Gatel, S. Joulie, F. Houdellier, C. Magen and M. J. Hytch, *Ultramicroscopy*, 2012, **115**, 78.
61. D. M. Eigler and E. K. Schweizer, *Nature*, 1990, **344**, 524.
62. J. A. Stroscio and D. M. Eigler, *Science*, 1991, **254**, 1319.
63. S. W. Hla, *J. Vac. Sci. Technol. B*, 2005, **23**, 1351.
64. O. Custance, R. Perez and S. Morita, *Nature Nanotechnology*, 2009, **4**, 803.
65. Ci, L.; Xu, Z.; Wang, L.; Gao, W.; Ding, F.; Kelly, K. F.; Yakobson, B. I.; Ajayan, P. M. *Nano Res.* 2008, **1**, 116.
66. F. Ming, K. Wang, S. Pan, J. Liu, X. Zhang, J. Yang and X. Xiao, *ACS Nano*, 2011, **5**, 7608.
67. M. Bieri, S. Blankenburg, M. Kivala, C. A. Pignedoli, P. Ruffieux, K. Müllen and R. Fasel, *Chem. Commun.*, 2011, **47**, 10239.
68. L. Cardenas, R. Gutzler, J. Lipton-Duffin, C. Fu, J. L. Brusso, L. E. Dinca, M. Vondráček, Y. Fagot-Revurat, D. Malterre, F. Rosei and D. F. Perepichka, *Chem. Sci.*, 2013,**4**, 3263
69. Vl. V. Voevodin, S. A. Zhumatiy, S. I. Sobolev, A. S. Antonov, P. A. Bryzgalov, D. A. Nikitenko, K. S. Stefanov and Vad. V. Voevodin, *Open Systems Journal*, 2012, **20**, 7.
70. http://computing.kiae.ru/

# Supplementary Information

*Formation of nickel-carbon heterofullerenes under electron irradiation*

A. S. Sinitsa, I. V. Lebedeva, A. A. Knizhnik, A. M. Popov, S. T. Skowron and E. Bichoutskaia

**Modified potential**

The same as in the original Brenner potential,[1] the energy of the system is represented as[2]

$$E_\mathrm{b} = \sum_i \sum_{j(>i)} E_{ij}, \qquad (S1)$$

where the energy $E_{ij}$ of the bond between atoms $i$ and $j$ separated by the distance $r_{ij}$ is given by the sum of repulsive and attractive terms

$$E_{ij} = V_\mathrm{R}(r_{ij}) - \bar{b}_{ij} V_\mathrm{A}(r_{ij}). \qquad (S2)$$

The repulsive interaction is determined by the two-body function

$$V_\mathrm{R}(r_{ij}) = f_{ij}(r_{ij}) A_{ij} \exp(-\lambda_{1,ij} r_{ij}), \qquad (S3)$$

where the cut-off function $f_{ij}(r)$ has the form

$$f_{ij}(r) = \begin{cases} 1, & r < R_{ij}^{(1)} \\ \dfrac{1}{2}\left[1 + \cos\left[\dfrac{\pi(r - R_{ij}^{(1)})}{(R_{ij}^{(2)} - R_{ij}^{(1)})}\right]\right], & R_{ij}^{(1)} \le r \le R_{ij}^{(2)} \\ 0, & r > R_{ij}^{(2)} \end{cases} \qquad (S4)$$

The attractive interaction is described by the two-body function

$$V_\mathrm{A}(r_{ij}) = f_{ij}(r_{ij}) B_{ij} \exp(-\lambda_{2,ij} r_{ij}) \qquad (S5)$$

multiplied by the function $\bar{b}_{ij}$ which describes the dependence of the interaction energy on the local coordination. The empirical bond order function $\bar{b}_{ij}$ is given by the sum of the average of the terms $b_{ij}$ and $b_{ji}$ corresponding to each atom in the bond and of the additional correction function $F_{ij}$, which is used to account for conjugated versus non-conjugated bonding and to avoid the overlapping of radicals,

$$\bar{b}_{ij} = (b_{ij} + b_{ji})/2 + F_{ij}(N_{ij}^\mathrm{C}, N_{ji}^\mathrm{C}, N_{ij}^\mathrm{conj})/2, \qquad (S6)$$

where $N_{ij}^\mathrm{C}$ is the number of carbon atoms bonded to atom $i$ in addition to atom $j$ and $N_{ij}^\mathrm{conj}$ is used to determine whether the bond between atoms $i$ and $j$ is a part of a conjugated system. The function $F_{ij}$ is non-zero only for bonds between two carbon atoms.

The bond order function $b_{ij}$ for each atom in the bond is determined by

$$b_{ij} = \left[1 + \sum_{k(\neq i,j)} G_{ijk}(\theta_{ijk}) f_{ik}(r_{ik}) \exp\left[\alpha_{ijk}\left((r_{ij} - R_{ij}^{(e)}) - (r_{ik} - R_{ik}^{(e)})\right)\right] + H_{ij}(N_i^C, N_i^{Ni})\right]^{-\delta}, \quad (S7)$$

where $R_{ij}^{(e)}$ is the equilibrium distance between atoms $i$ and $j$, $\theta_{ijk}$ is the angle between the bonds between atoms $i$ and $j$ and atoms $i$ and $k$ and $\delta$ is taken equal to 0.5 for all atoms. The function $H$ in this expression depending on the numbers $N_{ij}^{Ni}$ and $N_{ij}^C$ of nickel and carbon neighbors of atom $i$ in addition to atom $j$ was present also in the original Brenner potential[1] but it was taken equal to zero for the previous version of the Ni-C potential.[2] Here non-zero values of this function are introduced to improve description of low-coordinated nickel atoms.

The function $G_{ijk}(\theta)$ is taken in the form

$$G_{ijk}(\theta) = a_{ijk}\left[1 + \frac{c_{ijk}^2}{d_{ijk}^2} - \frac{c_{ijk}^2}{d_{ijk}^2 + (1 + \cos\theta)^2}\right]. \quad (S8)$$

As opposed to the original Brenner potential,[1] we assume that the parameters of the function $G_{ijk}(\theta)$ $a_{ijk}$, $c_{ijk}$ and $d_{ijk}$ depend on types of all three atoms $i$, $j$ and $k$.

The numbers $N_{ij}^C$ and $N_{ij}^{conj}$ are found as

$$N_{ij}^C = \sum_{C\,k(\neq j)} f_{ik}(r_{ik}), \quad (S9)$$

$$N_{ij}^{conj} = 1 + \sum_{C\,k(\neq i,j)} f_{ik}(r_{ik}) F_0(N_{ki}^C) + \sum_{C\,l(\neq i,j)} f_{jl}(r_{jl}) F_0(N_{lj}^C), \quad (S10)$$

where

$$F_0(x) = \begin{cases} 1, & x \leq 2 \\ [1 + \cos(\pi(x-2))]/2, & 2 < x < 3 \\ 0, & x \geq 3 \end{cases} \quad (S11)$$

In the modified Ni-C potential, the values of the function $F_{ij}$ are corrected according the original Brenner potential. In addition, $F_{CC}(1,2,2) = F_{CC}(2,1,2)$ are changed to $-0.0630$ to improve the graphene edge energies. The correct energies of carbon structures on the Ni (111) surface are kept by adjusting the parameters $a_{CCNi}$ and $a_{CNiC}$. The parameters of the modified potential are given in Table S1, Table S2 and Table S3. The parameters that are changed as compared to the previous version[2] are shown in bold. The function $H$ now takes non-zero values only in the case of $H = H_{NiC}(N^C, N^{Ni})$ for $N^{Ni} \leq 5$ and $N^C \geq 1$. Namely, $H_{NiC}(N^C \geq 1, N^{Ni} \leq 3) = 1.3$ and $H_{NiC}(N^C \geq 1, N^{Ni} = 4,5) = 2.5$.

**Table S1.** Two-body parameters of the potential.

| Parameters | C-C | C-Ni | Ni-Ni |
|---|---|---|---|
| $A$ (eV) | 2606 | 1866 | 1473 |
| $B$ (eV) | 1397 | 184.6 | 61.24 |
| $\lambda_1$ (Å$^{-1}$) | 3.2803 | 3.6768 | 3.2397 |
| $\lambda_2$ (Å$^{-1}$) | 2.6888 | 1.8384 | 1.2608 |
| $R^{(1)}$ (Å) | 1.7 | 2.2 | 3.0 |
| $R^{(2)}$ (Å) | 2.0 | 2.5 | 3.3 |
| $R^{(e)}$ (Å) | 1.3900 | 1.6345 | 2.0839 |

**Table S2.** Three-body parameters of the modified potential.

| Parameters | CCC | **CCNi** | **CNiC** | CNiNi | NiNiNi | NiNiC | NiCNi | NiCC |
|---|---|---|---|---|---|---|---|---|
| $\alpha$ (Å$^{-1}$) | 0 | 0 | 0 | 0 | 4.40 | 0 | 4.01 | 0 |
| $a$ | $2.08 \cdot 10^{-4}$ | **0.115** | **0.602** | $3.29 \cdot 10^{-3}$ | $9.28 \cdot 10^{-2}$ | 0 | $1.86 \cdot 10^{-4}$ | $1.22 \cdot 10^{-5}$ |
| $c$ | 330 | 0 | 0 | 5.72 | 7760 | 0 | 7410 | 240 |
| $d$ | 3.50 | 1.00 | 1.00 | 0.348 | 69.0 | 1.00 | 7.75 | 1.00 |

**Table S3.** Values of function $F_{CC}(i,j,k)$ for integer values of $i, j$ and $k$. Between integer values of $i, j$ and $k$, the function is interpolated by a cubic spline. All parameters not given are equal to zero, $F_{CC}(i,j,k>2) = F_{CC}(i,j,k)$.

| | $F_{CC}$ |
|---|---|
| (0,1,1), **(1,0,1)** | **0.0996** |
| (0,2,1), **(2,0,1)** | **0.0427** |
| (0,2,2), **(2,0,2)** | **-0.0269** |
| (0,3,1), **(3,0,1)** | **-0.0904** |
| (0,3,2), **(3,0,2)** | **-0.0904** |
| (1,1,1) | 0.1264 |
| (1,1,2) | 0.0108 |
| (1,2,1), **(2,1,1)** | **0.0120** |
| (1,2,2), **(2,1,1)** | **-0.0630** |
| (1,3,1), **(3,1,1)** | **-0.0903** |
| (1,3,2), **(3,1,2)** | **-0.0903** |
| (2,2,1) | 0.0605 |
| **(2,3,1), (3,2,1)** | **-0.0363** |
| **(2,3,2), (3,2,2)** | **-0.0363** |

**REFERENCES**

1. D. W. Brenner, *Phys. Rev. B*, 1990, **42**, 9458.

2. I. V. Lebedeva, A. A. Knizhnik, A. M. Popov and B. V. Potapkin, *J. Phys. Chem. C,* 2012, **116**, 6572.